\documentclass[11pt]{article}

\usepackage{graphicx,enumitem}
\usepackage{graphicx,psfrag,epsf}
{%
\centering\addtocounter{figure}{1}% if caption at bottom
\begin{enumerate}[%
itemsep=2pt,parsep=0em,
label={(\alph*)},
ref={\thefigure.(\alph*)}
]}%
{\end{enumerate}\addtocounter{figure}{-1}}

\usepackage[colorlinks=true,linkcolor=blue]{hyperref}%
\usepackage{graphicx}
\hypersetup{urlcolor=blue, citecolor=blue, colorlinks=true, linkcolor=blue} 
 
\usepackage[margin=1in]{geometry}

\usepackage[utf8]{inputenc}
\usepackage{amsmath}
\usepackage[english]{babel}
\usepackage{amssymb}
\usepackage{amsfonts}
\usepackage{bm}
\usepackage{graphicx}
\usepackage{amsmath}
\usepackage{amsthm}
\usepackage{xcolor}
\usepackage{parskip}
\usepackage{multicol}
\usepackage{float}

\usepackage{verbatim}

\floatstyle{plaintop}
\restylefloat{table}
\usepackage[tableposition=top]{caption}

\usepackage{authblk}

\usepackage{colortbl}

 \theoremstyle{definition}

\newtheorem{remark}{Remark}[section]

\usepackage{booktabs}
\usepackage{animate}

\usepackage{natbib}
\hypersetup{urlcolor=blue, citecolor=blue, colorlinks=true, linkcolor=blue} 

\numberwithin{equation}{section}
\numberwithin{figure}{section}
\numberwithin{table}{section}

\usepackage{blindtext}
\usepackage{lineno}

\begin{document}

\def\spacingset#1{\renewcommand{\baselinestretch}%
{#1}\small\normalsize} \spacingset{1}

\title{{\bf Neural Networks for Parameter Estimation in Geometrically Anisotropic Geostatistical Models}}

\author[1]{Alejandro Villazón}
\author[1]{Alfredo Alegr\'ia}
\affil[1]{Departamento de Matem\'atica, Universidad T{\'e}cnica Federico Santa Mar{\'i}a, Valparaiso, Chile}

\author[2,3]{Xavier Emery}
\affil[2]{Department of Mining Engineering\\ Universidad de Chile\\ Santiago, Chile}
\affil[3]{Advanced Mining Technology Center\\ Universidad de Chile\\ Santiago, Chile}

\maketitle

\begin{abstract} 
\noindent 
This article presents a neural network approach for estimating the covariance function of spatial Gaussian random fields {defined in a portion of the Euclidean plane}. Our proposal builds upon recent contributions, expanding from the purely isotropic setting to encompass geometrically anisotropic correlation structures, i.e., random fields with correlation ranges that vary across different directions. We conduct experiments with both simulated and real data to assess the performance of the methodology and to provide guidelines to practitioners.\\

\noindent {\emph Keywords}:  Deep learning; Mat\'ern covariance; Maximum likelihood;  {Statistical inference; Spatial random fields}; Variogram map
\end{abstract}

\section{Introduction}

Gaussian random fields constitute a valuable mathematical framework for analyzing geostatistical data across diverse fields of natural science, including climatology, environmental science, ecology, geosciences, among others \citep{chiles2012geostatistics}. This type of data is increasingly prevalent in the modern era, and the capacity to generate large volumes of data is ever-growing.

From a statistical perspective, maximum likelihood is the best alternative, in terms of efficiency and asymptotic properties, for estimating the first and second-order {moments} of a Gaussian random field, such as the mean, variance, and spatial correlation structure. However, the computational complexity of maximum likelihood scales cubically with the sample size due to the involvement of the determinant and inverse of the variance-covariance matrix in the likelihood function. This computational challenge has motivated researchers to explore and devise innovative methodologies and has gained substantial attention in recent literature, becoming a central point in ongoing research efforts. The significance of this topic is evident from recent competitions and publications addressing these critical issues; see \cite{heaton2019case}, \cite{huang2021competition}, and the references therein.

Deep learning techniques have emerged as a compelling alternative for performing estimation  in geostatistical models. Recently, \cite{gerber2021fast} explored a methodology based on neural networks to estimate the parameters in the covariance function of an isotropic random field. Fundamentally, the process of maximizing the likelihood function can be reconceptualized as mapping a vector of data (input) into an estimate (output). This mapping can be approximated by a neural network, provided a comprehensive training set is available. The problem is addressed through an extensive simulation of random fields using a wide range of parameter configurations, enabling the network to learn how to relate these elements.

The growing interest in this topic is reflected in additional recent contributions. For instance, \cite{sainsbury2023neural,sainsbury2023likelihood} employ different classes of neural networks to approximate Bayes estimators in a likelihood-free manner. Similarly, \cite{walchessen2023neural} use neural networks to learn the likelihood function in spatial models where the exact likelihood is intractable. \cite{lenzi2023neural} employ deep learning to estimate the parameters of max-stable models for which exact likelihood is challenging to evaluate. In a similar fashion, \cite{majumder2024modeling} investigate a methodology that combines deep learning and the Vecchia approximation, and its application in the study of extreme flooding events.  Methods specifically tailored for nonstationary models for spatial extremes have been developed by \cite{majumder2023deep}. On the other hand, \cite{gray2022use} address {the problem of predicting a spatial variable at unsampled locations} using a neural network approach. For a comprehensive review of the interface between spatiotemporal data and deep learning techniques, we refer readers to \cite{wikle2023statistical}.

%\textcolor{red}{XE: otra referencia que puede ser de inter\'es: \\https://www.sciencedirect.com/science/article/abs/pii/S0167947323000737}

This manuscript explores the application of neural networks for approximating maximum likelihood estimates, in the scenario of Gaussian random fields exhibiting a geometric anisotropy. In contrast to the isotropic case examined in previous literature, where spatial correlation ranges are the same in all directions, geometric anisotropy enables the modeling of phenomena with varying correlation ranges along different orientations. This proves valuable in many branches of knowledge. %For instance, it is useful for understanding pollution dispersion influenced by prevailing wind patterns (\textcolor{red}{referencia?}). 
For instance, it is useful for understanding climatologic \citep{Soderstrom1995}, epidemiologic \citep{Lecoustre1989}, or environmental \citep{Caetano2004} data influenced by prevailing wind patterns. 
Another example is the concentration of a metal in an ore deposit, which may exhibit an anisotropic spatial correlation due to major geological structures \citep{Samal2011} or to dominant mineralization directions \citep{Glacken2001}. 

Geometric anisotropy is modeled by deforming space through a positive definite matrix. In other words, the contour levels of an isotropic (radial) covariance function are transformed into ellipses (2D setting) or ellipsoids (3D setting). In terms of implementation, a significant difference from the isotropic case is that anisotropy requires, in the 2D setting, the addition of two extra parameters: one quantifies the principal angle of anisotropy, and the other measures the ratio between the semi-axes of the ellipse, which in some way indicates how distorted the space is. This necessitates developing neural networks with  different architectures and inputs to effectively handle these additional parameters. 

Since this problem inherently involves directionality, the architectures proposed in this work are designed to identify such patterns. Specifically, we propose two alternatives: the first one relies on a neural network that takes the realization of the random field as an input to provide parameter estimates, while the second variant considers the variogram map, which also clearly reflects the correlation structure across different directions.
We demonstrate the effectiveness of our proposals in two ways: first, by conducting a simulation experiment with conventional maximum likelihood as the benchmark, and second, by applying the methodologies to a real dataset of sea surface temperatures. Our numerical experiments are reproducible by accessing the repository \url{https://github.com/AlejandroVillazonG/AnisotropyEstimatorsNN}.
% \textcolor{red}{insertar link} \textcolor{red}{AV: Crearé otro repositorio sin historia y ordenado. Alguna sugerencia para el nombre??}. \textcolor{blue}{Quizás algo del estilo AnisotropyEstimatorsNN}

The article is organized as follows. Section \ref{background} contains preliminaries on Gaussian random fields with geometrically anisotropic correlation structures, and likelihood inference. Section \ref{NN} describes the estimation approaches based on neural networks and provides details about the networks architectures and training sets. In Section \ref{numerical}, we conduct numerical studies with both simulated and real data.
Section \ref{discussion} concludes the paper with a discussion.

\section{Geostatistical Model and Likelihood Inference}
\label{background}

Let $\{Z(\bm{s}):\bm{s}\in\mathcal{D}\}$ be a second-order stationary {Gaussian} random field, with $\mathcal{D}$ representing a spatial domain in the plane. Throughout, we assume that the random field has a zero mean (that is, the trend has been removed and we consider the residuals), and we focus on its covariance function. We work under the framework of geometric anisotropy, where the range of spatial dependence varies depending on the spatial orientation. Formally, the covariance function of the random field adopts the form
\begin{equation}
   \text{cov}\left(Z(\bm{s}),Z(\bm{s}+\bm{h}) \right) = \sigma^2 \varphi\left( \left[ \bm{h}^\top \bm{\Omega} \bm{h} \right]^{1/2}; \theta\right), \qquad \bm{s},\bm{s}+\bm{h}\in\mathcal{D},
\end{equation}
where $^\top$ denotes \emph{transpose} and $\varphi(\cdot; \theta)$ is a given valid (positive semi-definite) radial correlation function ($\varphi(0;\theta)=1$), with $\theta>0$ standing for a range parameter. Here, $\sigma^2 > 0$ is a parameter that regulates the variance of the random field, and $\bm{\Omega}$ is a positive definite matrix that deforms the contour levels of $\varphi(\cdot;\theta)$, mapping circles into ellipses, allowing for a correlation model with varying ranges across different directions. 

We parameterize the matrix $\bm{\Omega}$ in a conventional way \citep[p. 180]{journel1978},
$\bm{\Omega} := \bm{\Omega}(\alpha,\lambda) = \bm{P}(\alpha)^\top\bm{D}(\lambda) \bm{P}(\alpha)$, where
$$\bm{P}(\alpha) = \begin{bmatrix}
\cos(\alpha) & \sin(\alpha)\\
-\sin(\alpha) & \cos(\alpha)
\end{bmatrix} \quad \text{ and } \quad \bm{D}(\lambda) =  \begin{bmatrix}
    1 & 0\\
    0 & \lambda
\end{bmatrix}.$$
The parameter $\lambda \in (0,1)$ represents a ratio that allows us to evaluate the discrepancies, in terms of the rate of decay of the correlation, between the semi-axes of the ellipse, and $\alpha \in (0,\pi)$ is an angle of rotation. Note that we restrict the angle to such an interval to avoid identifiability issues, as $\alpha$ and $\alpha+\pi$ correspond to the same geometric anisotropy. 

The special case of zonal anisotropy, representing a degenerate scenario of geometric anisotropy when the ratio tends to zero (see \citealp{chiles2012geostatistics}), is not addressed in this study. This particular case warrants a dedicated analysis, and as such, it falls outside the scope of our current investigation.

A common choice for $\varphi(\cdot;\theta)$ is the Matérn model, a gold standard in spatial statistics, which takes the form \citep{stein1999interpolation}
$$\varphi(t;\theta) = \frac{2^{1-\nu}}{\Gamma(\nu)} \left(\frac{t}{\theta}\right)^\nu K_{\nu}\left(\frac{t}{\theta} \right), \qquad t\geq 0,$$
where $\nu>0$ is a shape parameter, $\Gamma$ is the gamma function, and $K_{\nu}$ is the modified Bessel function of the second kind. When $\nu = 1/2 + m$, where $m$ is a non-negative integer, this model simplifies to an exponential model multiplied by a polynomial of degree $m$, facilitating its implementation. The parameter $\nu$ regulates the degree of smoothness (mean square differentiability) of the sample paths of the associated random field.

The parameters involved in the covariance structure are typically estimated through maximum likelihood. Indeed, given a realization $Z(\bm{s}_1),\hdots,Z(\bm{s}_n)$, and using the notation $\bm{Z} = (Z(\bm{s}_1),\hdots,Z(\bm{s}_n))^\top$,
the log-likelihood function, up to a constant, is given by
\begin{equation}
    \label{loglik}
    \ell(\bm{\xi};\bm{Z}) = -\frac{1}{2}\left[ n\log(\sigma^2) + \log|\bm{R}(\alpha,\lambda,\theta)| +  \bm{Z}^\top \left[\sigma^2 \bm{R}(\alpha,\lambda,\theta)\right]^{-1}\bm{Z}  \right],
\end{equation}
where the $(i,j)$-th entry of the correlation matrix $\bm{R}(\alpha,\theta,\lambda)$ is  $\varphi\left(  \left[ (\bm{s}_i-\bm{s}_j)^\top \bm{\Omega} (\bm{s}_i-\bm{s}_j) \right]^{1/2}; \theta\right)$, and $\bm{\xi}=(\sigma^2,\alpha,\lambda,\theta)$ is the vector of parameters. The maximum likelihood estimate of $\bm{\xi}$ is given by $\widehat{\bm{\xi}}=\text{argmax}_{\bm{\xi}\in\bm{\Xi}} \ell(\bm{\xi};\bm{Z})$, where $\bm{\Xi}$ is the parameter space.

The determinant and the inverse pose greater challenges when evaluating (\ref{loglik}). Cholesky factorization is the most common approach for performing these tasks; however, it has a computational complexity of $O(n^3)$ \citep{aho1974design}, making its implementation computationally demanding when the sample size is large.

Throughout this analysis, we focus on the parameters $\alpha$, $\lambda$, and $\theta$. It is worth noting that profile likelihood estimation could be employed for $\sigma^2$, allowing us to concentrate on the key parameters involved in the correlation structure.

\section{Neural Network Approach for Estimation}
\label{NN}

As previously described, maximum likelihood can be conceptualized as a mapping, $\mathcal{F}_{\text{ML}}: \mathbb{R}^n \rightarrow \bm{\Xi}$, which takes an $n$-dimensional vector of spatial data $\bm{Z}$ and maps it to $\widehat{\bm{\xi}}$. Neural networks can emulate this mapping when trained on multiple vectors $\bm{Z}$ corresponding to different parameter values $\bm{\xi}$. By conducting numerous simulations across various parametric settings, we aim to learn this mapping, directly associating input data with target parameters. This approach results in an appealing tool for subsequent inferences, effectively bypassing the traditional likelihood paradigm. Let us now highlight crucial aspects involved in the implementation of this methodology.

\subsection*{Model Architecture} 
The specific form of the neural network is termed a model architecture and is expressed as a sequence of layers. In our case, we leverage a combination of dense and convolutional layers with linear and ReLU activation functions, providing the flexibility needed for the complexity of the spatial random fields under consideration. For a thorough exploration of this topic, we recommend consulting more detailed references \citep{sharma2017activation, gu2018recent}.

\subsection*{Training the Neural Networks}
\begin{figure}
    \centering
    \includegraphics[scale=0.072]{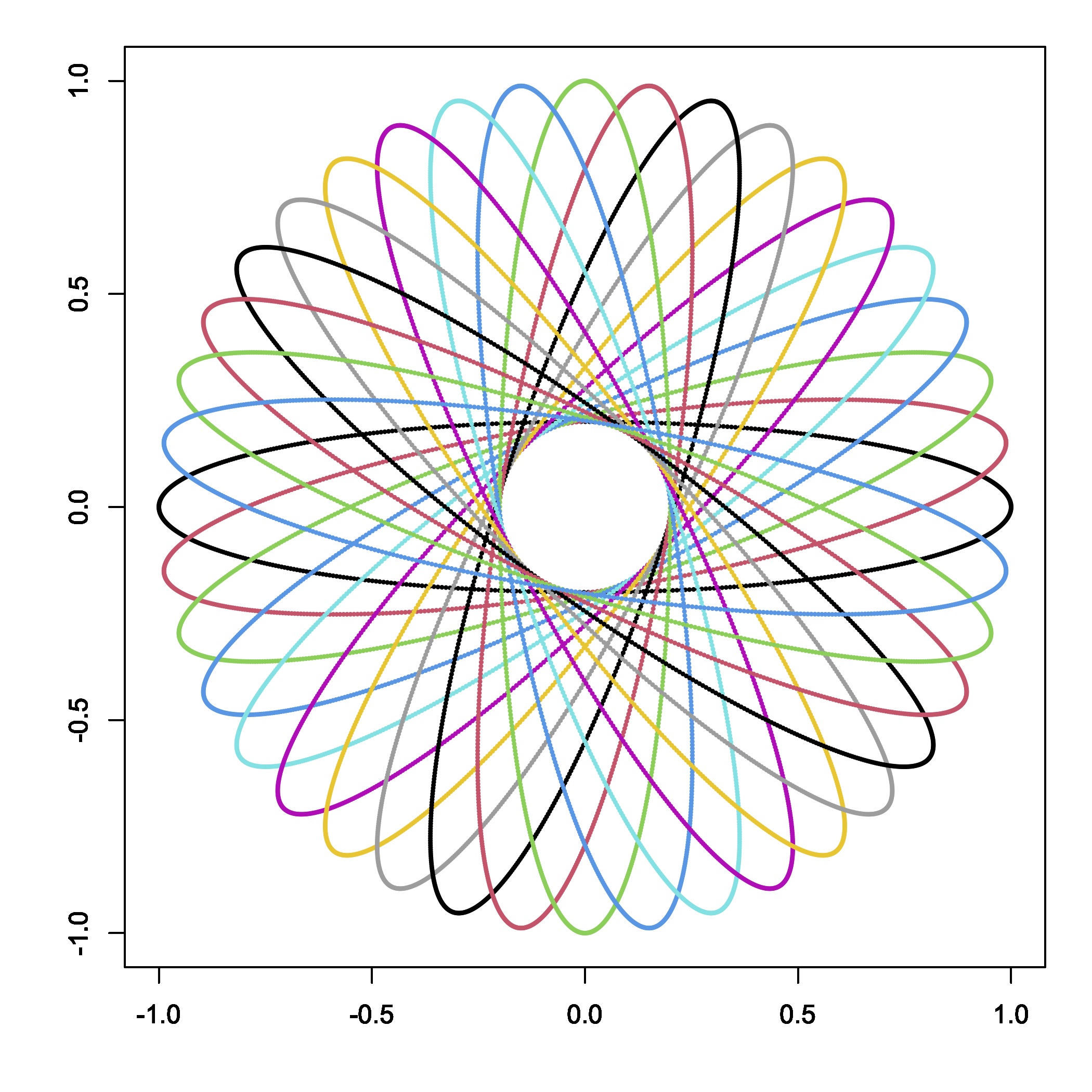}
        \includegraphics[scale=0.072]{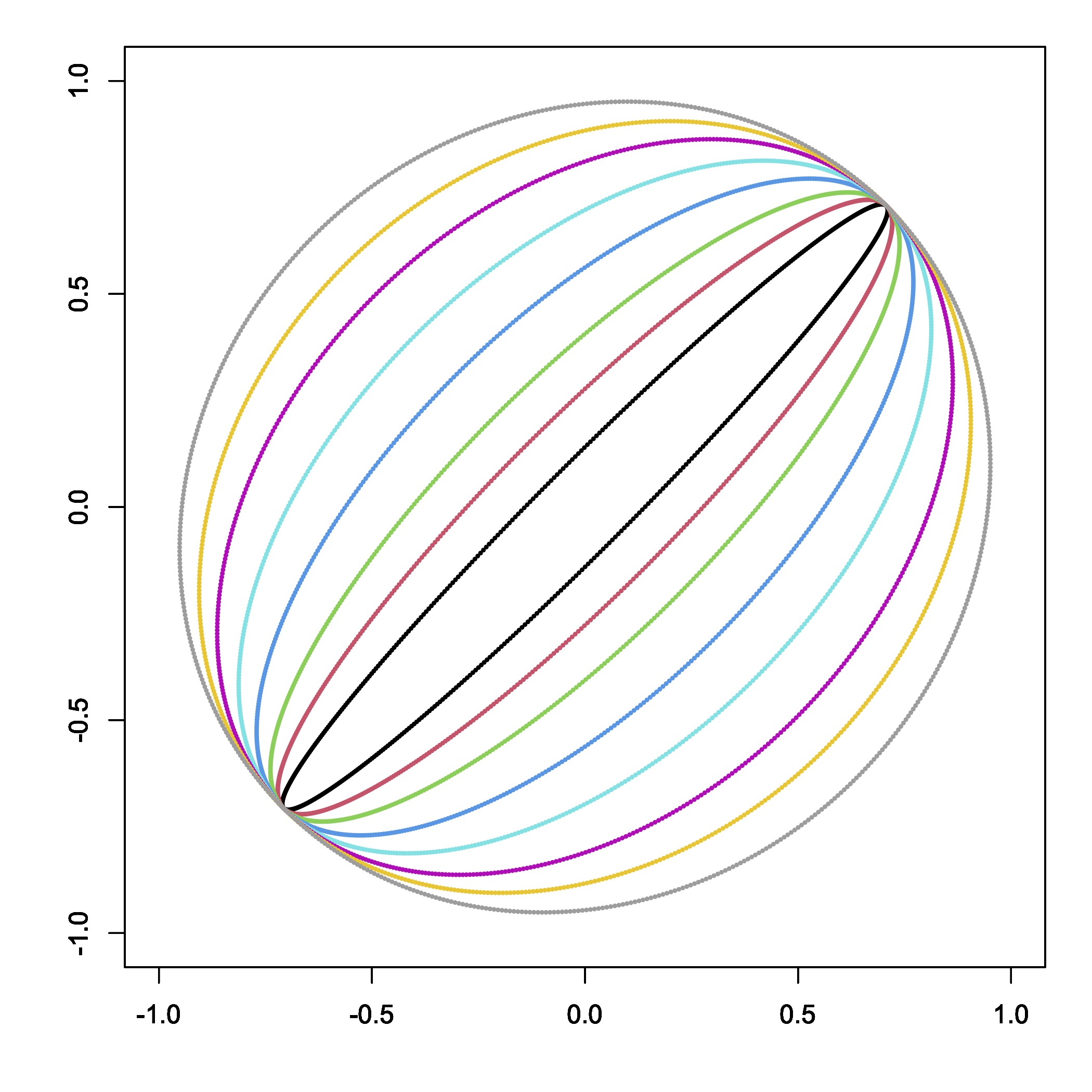}
    \includegraphics[scale=0.072]{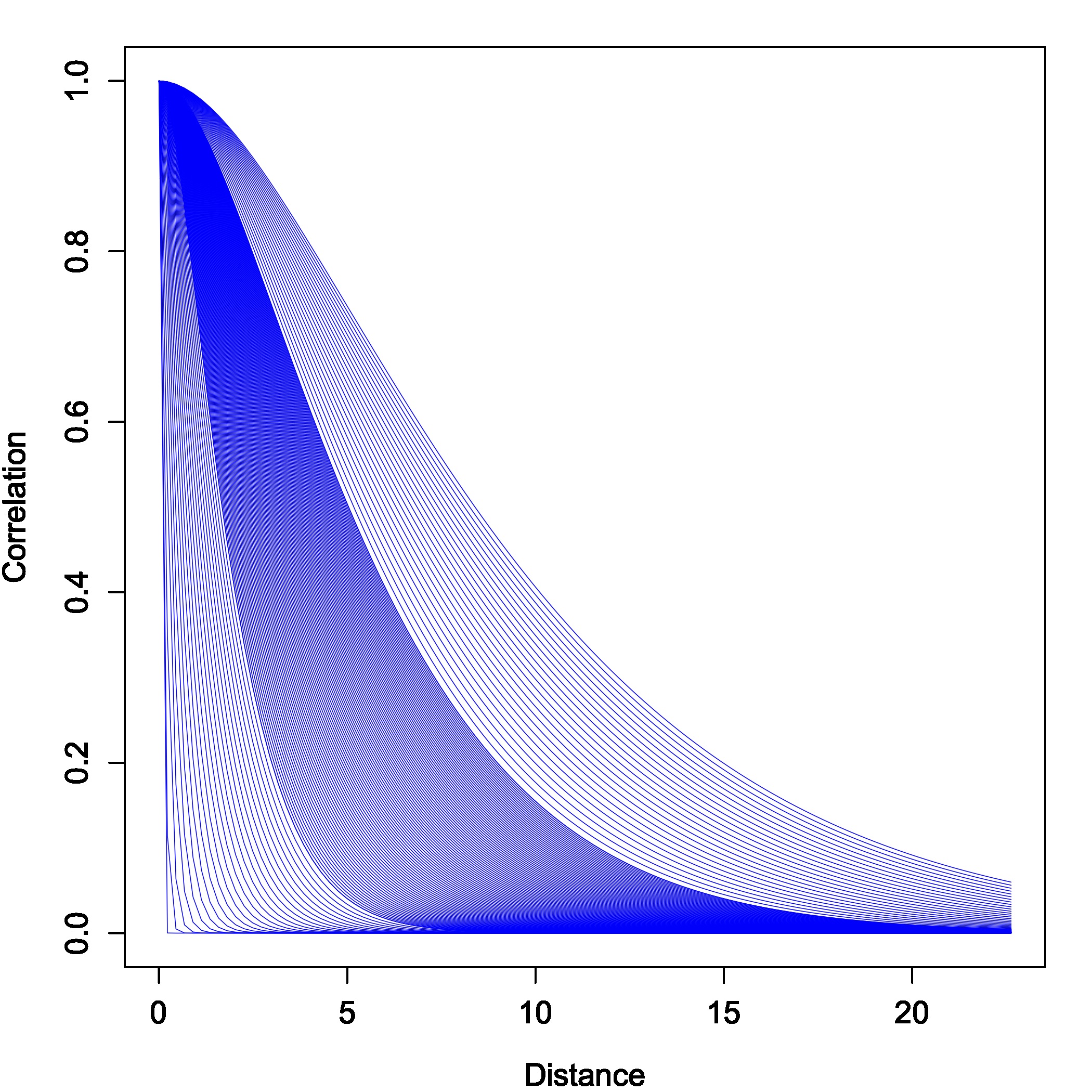} 
    \caption{(Left) Directions considered in our study (with a fixed ratio $\lambda=0.2$). (Middle) Some of the ratios considered in our study (with a fixed direction $\alpha=\pi/4$); we only display $8$ ellipses to ease visualization. (Right) Correlation functions $t\mapsto \varphi(t;\theta)$ for the ranges considered in our study. }
    \label{fig:training}
\end{figure}

 The training step involves generating training data, defining a loss function, and configuring an optimizer. 
    To cover the parameter space, we adopt the following scheme. The angular parameter $\alpha$ is discretized into $20$ equidistant points, ranging from $0$ to $\pi$ radians, excluding $\pi$, capturing a broad spectrum of orientations. Figure \ref{fig:training} (left panel) illustrates the considered orientations of the ellipses for a fixed ratio of $0.2$. Similarly, the ratio parameter $\lambda$ is divided into three segments: $(0, 0.3)$, $[0.3, 0.7)$, and $[0.7, 1]$, with $25$, $100$, and $25$ equidistant points in each segment, respectively. This stratification ensures a comprehensive exploration of different anisotropy intensities (Figure \ref{fig:training} (middle panel) shows the effect of different ratios; we only report 8 ratios to facilitate visualization). For the range parameter $\theta$ we consider three intervals: $[0.02, 1)$, $[1, 3)$, and $[3, 5]$, with $25$, $100$, and $25$ equidistant points in each interval, respectively. This design allows for a thorough examination of both short-range and long-range spatial dependencies, considering the spatial domain $\mathcal{D}=\{1,\hdots,16\}^2$ and a Mat\'ern correlation model (with smoothness parameter $\nu=1.5$) as a reference. More precisely, the practical range, i.e., the distance at which the correlation is lower than $0.05$, varies from $0.095$ to $23.8$. Figure \ref{fig:training} (right panel) shows the landscape of correlation functions $\varphi(\cdot;\theta)$ so obtained.

In total, our parameter grid comprises $20 \times 150 \times 150 = 450,000$ configurations, each representing a unique combination of $\alpha$, $\lambda$, and $\theta$. The non-uniformity of the grid is motivated by the necessity to emphasize certain regions of interest within the parameter space; for example, extremely small or large values of $\theta$ are less represented in the training set due to their lesser relevance to the spatial domain considered. With the efficient vectorized implementation using Python's NumPy and Google Colab's GPU back-end, we simulated $450,000$ realizations with the aforementioned features in approximately $36$ minutes. 

%\textcolor{red}{(una simulación espectral permitiría que las realizaciones cambien poco cuando un parámetro cambia poco. No sé qué método usaron y si este método tiene esta buena propiedad.)}
 
We employ the mean absolute error (MAE) loss function to quantify the accuracy of the model on the training samples. Given the varied scales of parameters $\alpha$, $\lambda$, and $\theta$, we scale them by subtracting their mean and dividing by their standard deviation. This normalization ensures that each parameter contributes equally to the optimization process. The AdamW optimizer, a state-of-the-art optimizer, is utilized with default parameters, including a learning rate of $0.01$. This optimizer is well-suited for efficiently updating weights and achieving convergence in the training process.

  \subsection*{Neural Network Architectures}

We consider two variants of this methodology: 

\begin{itemize}
    \item  The first one, denoted by the acronym NF, employs complete random fields realizations as input.  The NF model architecture includes a combination of dense and convolutional layers, as well as linear and ReLU activation functions, which is detailed in Table \ref{tab:arqNF}. The choice of convolutional layers is motivated by their effectiveness in processing image-like structures, treating the simulated random fields as images. Given the geometrically anisotropic nature of the images, we augment the training data by rotating the images by $180$ degrees, effectively doubling the number of fields learned in each epoch to $900,000$. The optimizer updates the weights every $500$ samples, known as the batch size, with each epoch lasting approximately $47$ seconds. {The loss level decreases and stabilizes} after $50$ epochs, resulting in a total training time of $39$ minutes.

% \textcolor{red}{(qué significa ``satisfactory''? suena muy cualitativo)}
%\textcolor{blue}{quizás Alejandro puede precisar "aceptable", quizás hay una opción por default, aclarar eso}

\begin{table}
    \centering
    \begin{tabular}{lrrrlr}
        \toprule
        layer type & output shape & filters & kernel size & activation & parameters \\
        \midrule
        2D convolution &  [–, 8, 8, 128] &  128 & 9×9 & ReLU & 10,496 \\
        2D convolution &  [–, 4, 4, 256] &  256 & 5×5 & ReLU & 819,456 \\
        2D convolution &  [–, 1, 1, 512] &  512 & 4×4 & ReLU & 2,097,664 \\
        2D convolution &  [–, 1, 1, 1024] &  1024 & 1×1 & ReLU & 525,312 \\
        flatten & [–, 1024] & & & & 0\\
        dense & [–, 300] & & & ReLU & 307,500 \\
        dense & [–, 3] & & & linear & 903 \\
        \midrule
        \multicolumn{4}{l}{total trainable parameters:} & \multicolumn{2}{r}{3,761,331} \\
        \bottomrule
    \end{tabular}
    \caption{Summary of the NF model.  It is a sequential model taking input of shape [–, 16, 16, 1] and mapping it to three scalar values of shape [–, 3]. After the training, the outputs become estimates of $\alpha$, $\lambda$, and $\theta$.}
    \label{tab:arqNF}
\end{table}

\item The second one, denoted by the acronym NV, processes variogram maps of size $13\times 13$. This version offers flexibility, in comparison to NF, as it allows for handling realizations with missing values and irregularly spaced locations, focusing solely on the posterior variogram. The architecture is similar to NF, with some hyperparameter adjustments detailed in Table \ref{tab:arqNV}. While calculating variogram maps is computationally expensive, taking $1$ hour, the training of the NV model is efficient, taking around $5$ seconds per epoch with $100$ epochs being sufficient for optimal performance, leading to a total training time of approximately $9$ minutes.

\begin{table}
    \centering
    \begin{tabular}{lrrrlr}
        \toprule
        layer type & output shape & filters & kernel size & activation & parameters \\
        \midrule
        2D convolution &  [–, 6, 6, 128] &  128 & 8×8 & ReLU & 8,320 \\
        2D convolution &  [–, 3, 3, 128] &  128 & 4×4 & ReLU & 262,272 \\
        2D convolution &  [–, 1, 1, 256] &  256 & 3×3 & ReLU & 295,168 \\
        2D convolution &  [–, 1, 1, 512] &  512 & 1×1 & ReLU & 131,584 \\
        flatten & [–, 512] & & & & 0\\
        dense & [–, 300] & & & ReLU & 153,900 \\
        dense & [–, 3] & & & linear & 903 \\
        \midrule
        \multicolumn{4}{l}{total trainable parameters:} & \multicolumn{2}{r}{852,147} \\
        \bottomrule
    \end{tabular}
    \caption{Summary of the NV model. It is a sequential model taking input of shape [–, 13, 13, 1] and mapping it to three scalar values of shape [–, 3]. After the training, the outputs become estimates of $\alpha$, $\lambda$, and $\theta$.}
    \label{tab:arqNV}
\end{table}

\end{itemize}

\begin{figure}
    \centering
    \includegraphics[scale=0.1]{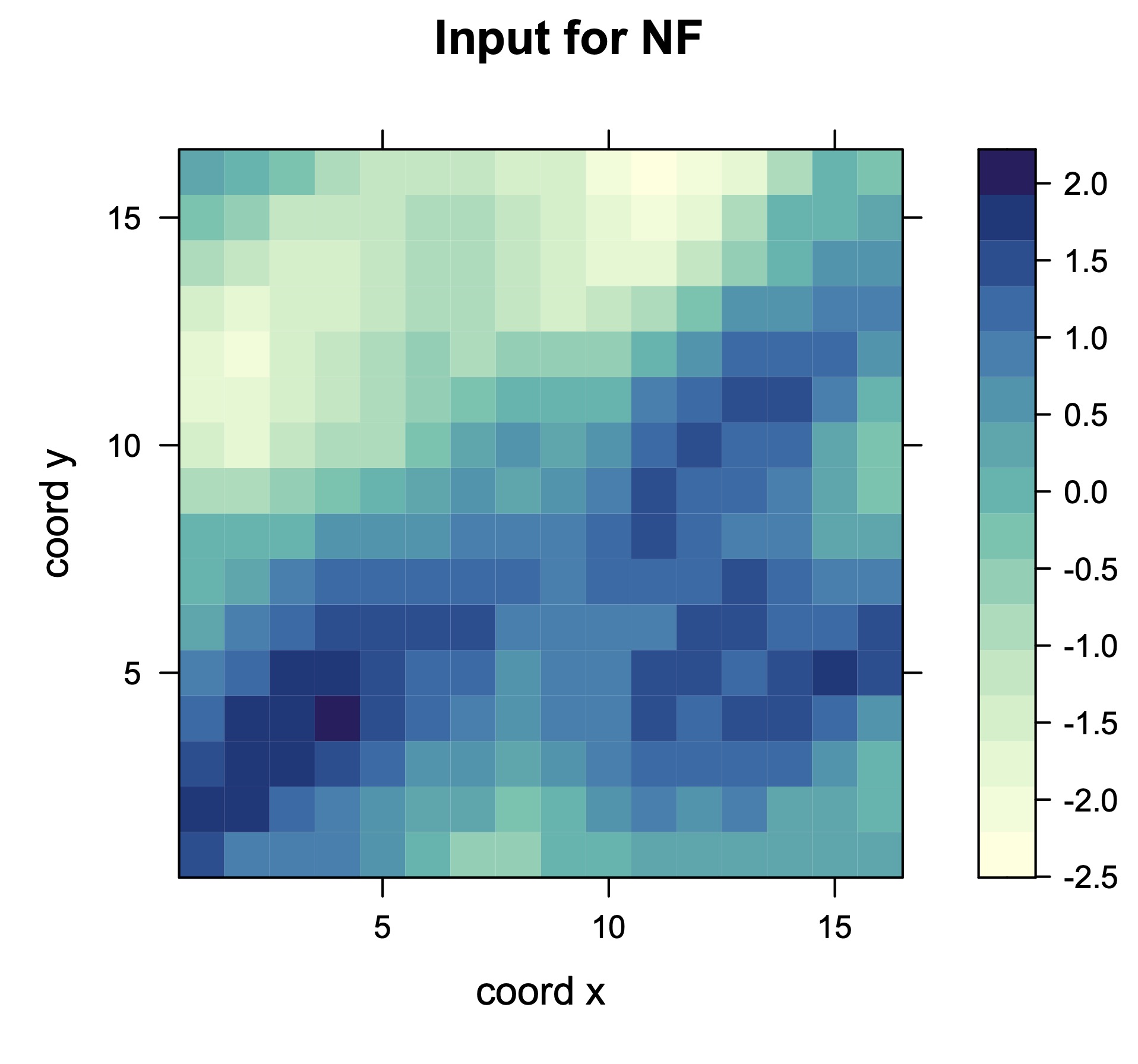}
    \includegraphics[scale=0.1]{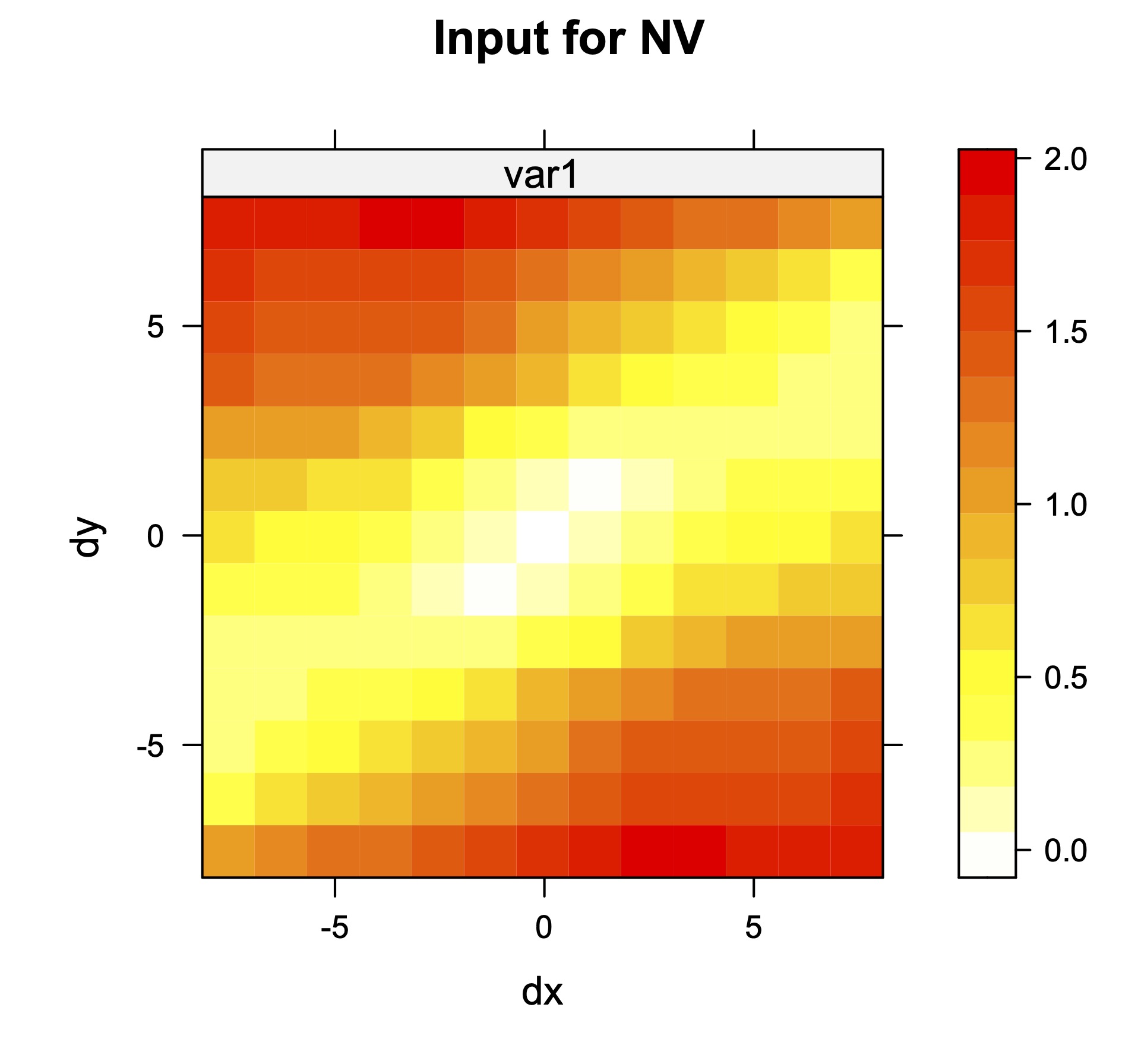}
    \caption{Inputs for NF and NV methodologies. The input for NF is a realization of a {Gaussian} random field over $256$ regularly spaced locations, while the input for NV is the variogram map with $13 \times 13$ pixels. In both cases, the output (vector of parameter values) is $(\alpha,\lambda,\theta) = (\pi/4,0.3,2)$.}
    \label{fig:inputs}
\end{figure}

Figure \ref{fig:inputs} illustrates the inputs associated with each version of this methodology. On the left, the input for NF is an image (a realization of a {Gaussian} random field) with 256 pixels, while on the right, the input for NV is an image (the variogram map of the realization) with 169 pixels. For both cases, the output (i.e., the vector of parameter values) is $(\alpha,\lambda,\theta) = (\pi/4,0.3,2)$.

%\textcolor{red}{AV: Para NF mantuvimos la estructura de la arquitectura del caso isotrópico, combinación de capas convolucionales y densas, la diferencia radica en la elección de filtros y tamaño de kernel, buscamos aquellos que se ajustaran a nuestro problema. Para NV la estructura de la red cambia totalmente al caso isotrópico, ya que el variograma ahora es bidimensional, siendo necesario incluir capas convolucionales a las capas densas del caso isotrópico, la nueva arquitectura imita la estructura de NF con ajustes en la capa de entrada y elección específica de filtros y tamaños de kernel.}

\begin{remark}
Let us provide some remarks on the proposed methodologies. 
For NF, we retained the architecture from the isotropic case described in \cite{gerber2021fast}, which integrates convolutional and dense layers. The main variation lies in the selection of filters and kernel sizes, which we optimized to better address our specific problem. In contrast, for NV, the network structure differs significantly from the isotropic case due to the two-dimensional nature of the variogram. This requires incorporating additional convolutional layers alongside the dense layers used previously. The new architecture is designed to resemble the NF structure but with modifications to the input layer and tailored choices of filters and kernel sizes.
\end{remark}

\section{Numerical Results}
\label{numerical}

%\textcolor{red}{En ambos casos, sugiero también ajustar un ML isótropo y ver cuál es la verosimilitud alcanzada en cada método: ML anisótropo, ML isótropo, NF y NV anisótropos. Esto permitiría argumentar que introducir la anisotropía es algo relevante al aumentar (considerablemente?) la verosimilitud, y que sigue siendo factible sin introducir sobreajustes o parámetros arbitrarios/no inferibles.}

\begin{figure}
    \centering
\includegraphics[scale=0.08]{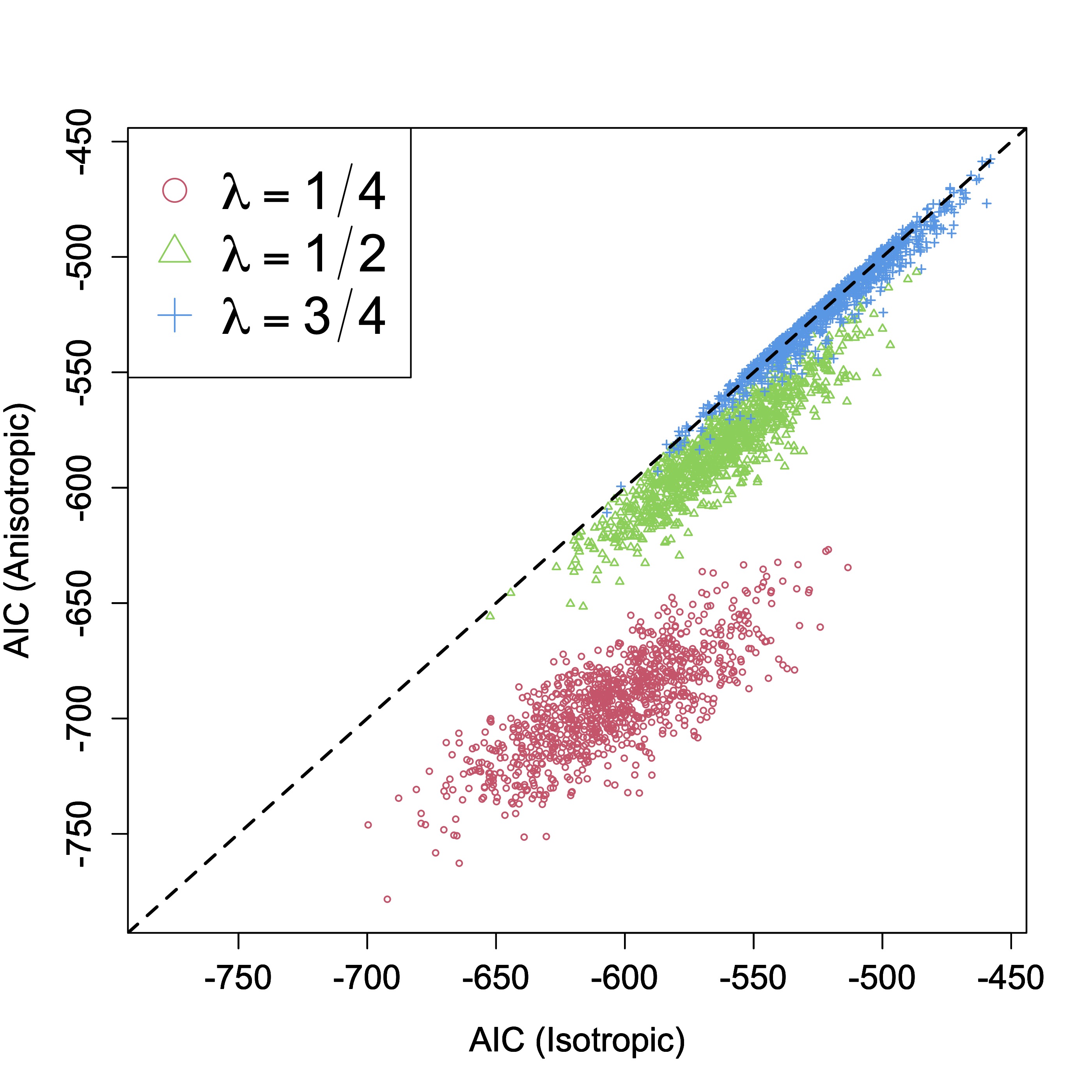}
    \caption{Scatter plots of Akaike Information Criterion (AIC) considering  isotropic (misspecified) versus anisotropic (correctly specified) models. The dashed line is the identity.}
    \label{fig:AICs}
\end{figure}

Before applying the tools introduced in this paper, we conduct a brief investigation to quantify the statistical improvements gained by considering anisotropic models in scenarios with inherent anisotropy. Specifically, we use the Mat\'ern model with parameters $\nu=3/2$, $\sigma^2=1$, $\theta=2$, $\alpha=\pi/4$, and three scenarios for the ratio $\lambda=1/4, 1/2, 3/4$, representing different levels of anisotropy intensity. For each scenario, we simulate $1,000$ independent Gaussian random fields on $\mathcal{D}=\{1, \hdots, 16\}^2$ and estimate the parameters through maximum likelihood, considering both isotropic (misspecified) and anisotropic (correctly specified) models. Figure \ref{fig:AICs} shows a scatter plot of the Akaike Information Criterion (AIC). Since lower AIC values indicate  better statistical performance, this experiment clearly demonstrates the improvement in goodness-of-fit achieved by employing a more sophisticated model with physically interpretable parameters. The improvement is particularly pronounced when $\lambda$ equals $1/4$ or $1/2$, as anticipated.

\subsection{Simulated Data}

We now conduct a simulation study to assess the performance of the neural network methodologies. We consider a set of $23,120$ parameter configurations to construct a validation set and evaluate the accuracy of the trained neural networks. Specifically, we focus on models with the angle $\alpha \in [0,\pi)$ and parameters $\lambda$ and $\theta$ within their dense training regions ($\lambda \in [0.3, 0.7)$ and $\theta \in [1,3)$). For testing purpose, we simulate $5$ random fields for each parameter configuration, maintaining the covariance structure considered above, resulting in a set of $115,600$ fields.

\begin{figure}
    \centering
   \includegraphics[width=0.9\textwidth]{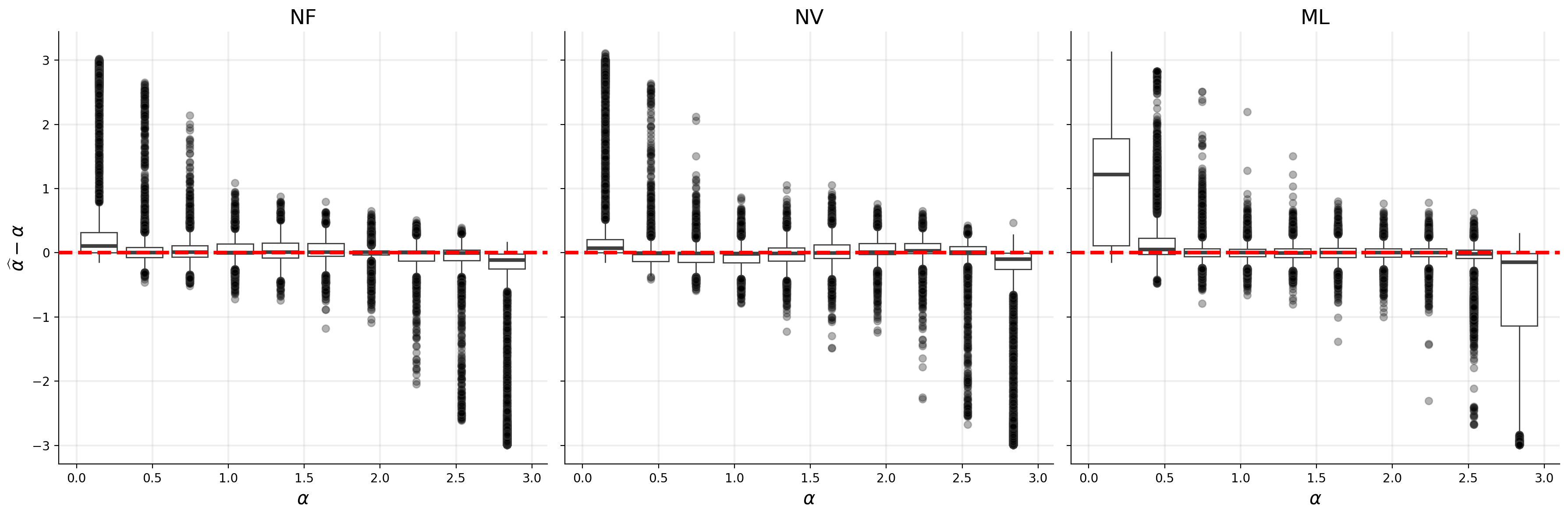}    
        \includegraphics[width=0.9\textwidth]{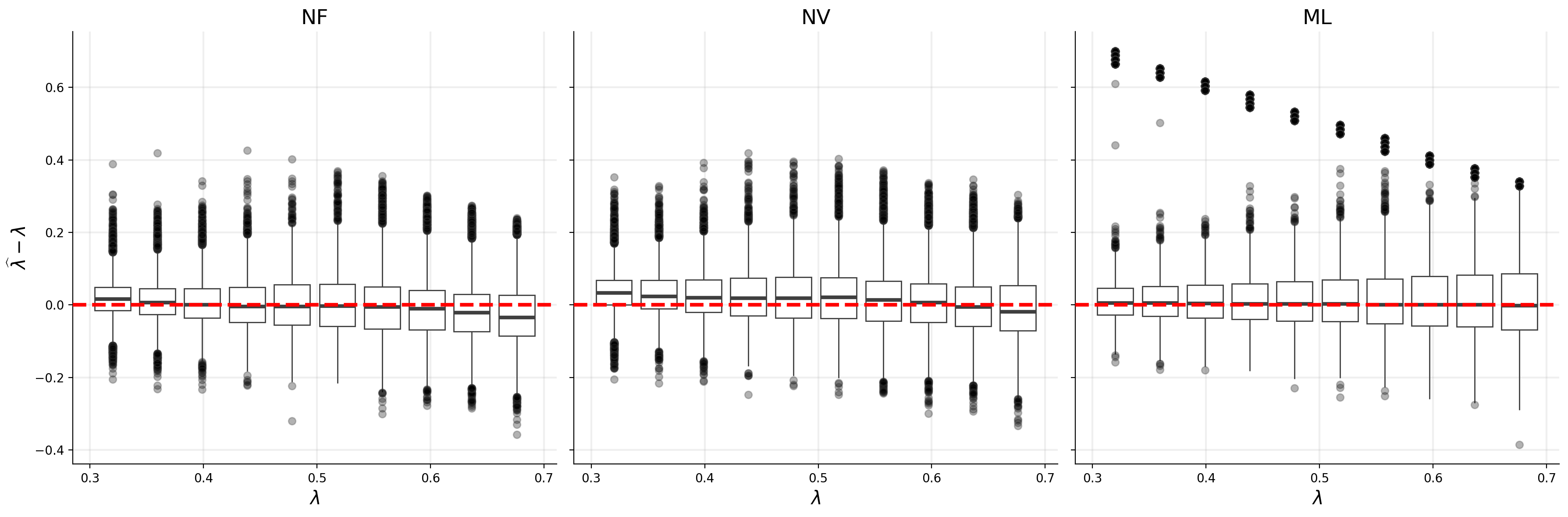}  
        \includegraphics[width=0.9\textwidth]{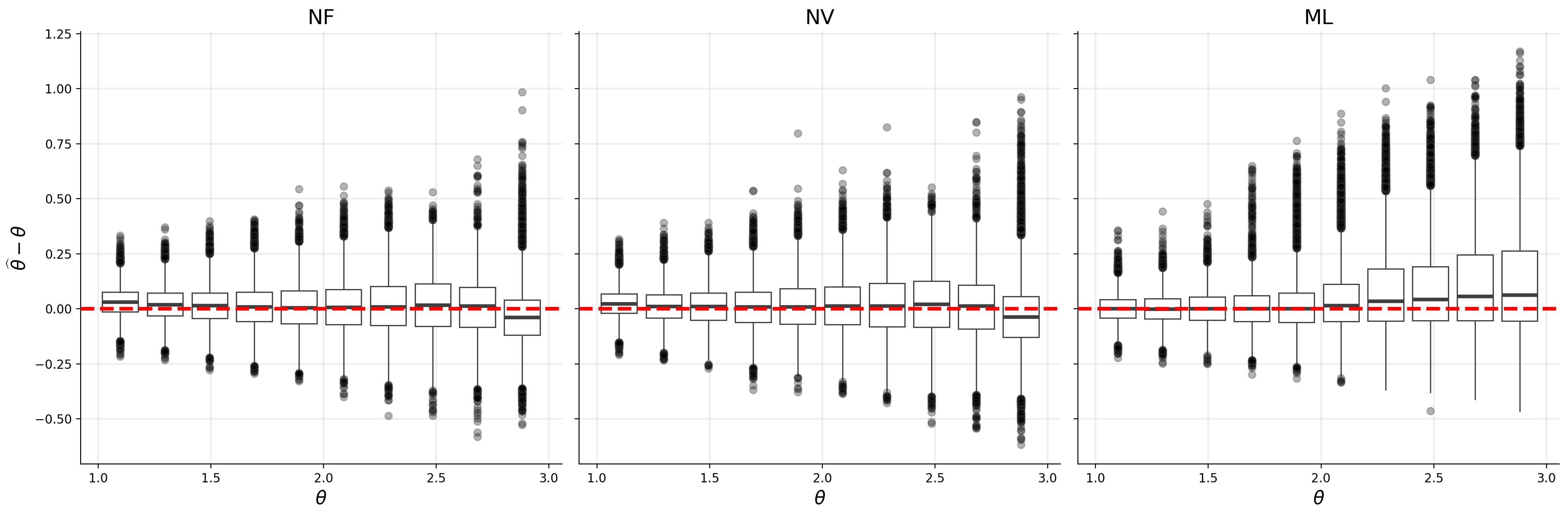}
    \caption{Centered boxplots for $\widehat{\alpha}$, $\widehat{\lambda}$ and $\widehat{\theta}$ estimates obtained by NF, NV and ML methods.}
    \label{fig:boxplots}
\end{figure}

\begin{figure}
    \centering   \includegraphics[width=0.9\linewidth]{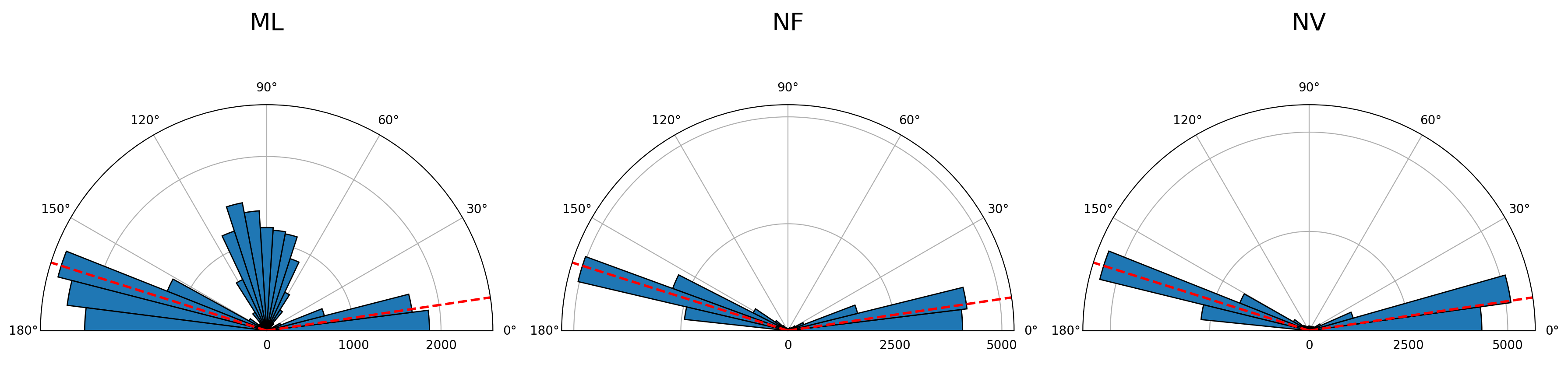}
    \caption{Rose diagrams of the $\alpha$ estimates for each method, when the true values of $\alpha$ (dashed red lines) are close to the extremes of the parameter space.}
    \label{fig:rose}
\end{figure}

In Figure \ref{fig:boxplots}, we compare the parameter estimates across the NF, NV, and ML methods.   Table \ref{tab:summary} also provides a comprehensive overview of the bias and standard deviation of parameter estimates for each method.  Although all methods seem to perform reasonably well in terms of bias, there are several key aspects to highlight:

\begin{itemize}
    \item For the parameter $\alpha$, all three methods show apparent inaccuracies at the extremes of the parameter space, {as indicated by biased boxplots and large standard deviations in Table \ref{tab:summary}}. However, this effect aligns with the nature of $\alpha$ corresponding to directionality, where these angles result in approximately the same direction. To gain a clearer understanding of this issue,  Figure \ref{fig:rose} presents rose diagrams of the estimates at these extreme values, highlighting the precision of NF and NV, {when viewed from a circular perspective}. In contrast, ML exhibits a notable number of estimates that significantly deviate from the true values.

\item Regarding the range parameter $\theta$, the performance of the ML method deteriorates, primarily in terms of variability, as $\theta$ increases. This finding is consistent with previous literature on the estimation of the range parameter for processes indexed by a fixed spatial domain \citep{zhang2004inconsistent}. Interestingly, the NF and NV methods, despite lacking explicit knowledge of the underlying statistical assumptions, demonstrate more robust performance in this regard.

\item The ML method consistently estimates $\lambda$ as $1$ for various test fields (the systematic outliers at the top of the corresponding boxplot), inaccurately implying an absence of anisotropy. Interestingly, the NF and NV methods do not exhibit this issue.

\end{itemize}
%\textcolor{red}{AA: Puedo ampliar el p\'arrafo anterior, señalando literatura donde ML presenta ciertas limitaciones, y aparentemente NN supera algunos de estos obst\'aculos}

%\textcolor{red}{AA: Planeo agregar un gráfico extra comparando la variabilidad de las estimaciones para los diferentes métodos y parámetros.}

%\textcolor{red}{AA: Alejandro, el título del primer panel, en los boxplots, debería ser "NF" en lugar de "NN"}

%\textcolor{red}{AA: Alejandro, podrías por favor hacer un "rose diagram" para las estimaciones del ángulo, por ejemplo, en R se puede hacer con el package "circular". Este diagrama es esencialmente un histograma pero en un círculo. Específicamente, considera NF, y toma todas las estimaciones del primer y último bin (es decir, ángulos que están en torno al ángulo 0). Con todas estas estimaciones, hace el diagrama de rosa. Luego, hacer lo mismo para NV y ML.}

Overall, the proposed neural network methods consistently demonstrate performance comparable to that of the traditional ML approach across the studied configurations. 

Regarding computation times, utilizing Google Colaboratory's GPU backend, the ML method took 10.5 hours to complete the estimation, while the trained neural network methods accomplished the same task in just 10 seconds. This significant reduction in computational burden positions neural network approaches as compelling alternatives, particularly when dealing with parameter estimation tasks in the presence of large datasets.

\begin{table}
    \centering
\begin{tabular}{llrrrrrrr} \hline \hline
\multicolumn{2}{l}{} & \multicolumn{3}{c}{bias} & & \multicolumn{3}{c}{standard deviation} \\ \cline{3-5} \cline{7-9}
& & NF & NV & ML  & & NF & NV & ML \\ \hline
$\alpha$  & 0.14751&1.1955&0.6161&\cellcolor{lightgray}{0.4676} & & 1.0239&1.0703&\cellcolor{lightgray}{0.9314}\\
& 0.4475&0.2595&0.0444&\cellcolor{lightgray}{-0.0129} & &  0.4927&0.2903&\cellcolor{lightgray}{0.1985}\\
& 0.746&\cellcolor{lightgray}{0.0160}&0.0225&-0.0347  & & 0.1548&0.1676&\cellcolor{lightgray}{0.1518}\\
& 1.0445&\cellcolor{lightgray}{0.0024}&0.0224&-0.0393 & & \cellcolor{lightgray}{0.1192}&0.1512&0.1572\\
& 1.343&\cellcolor{lightgray}{0.0012}&0.0223&-0.0180 & & \cellcolor{lightgray}{0.126}&0.1584&0.1590\\
& 1.6415&\cellcolor{lightgray}{-0.0003}&0.0175&0.0062 && \cellcolor{lightgray}{0.1303}&0.1521&0.1668\\
& 1.94&\cellcolor{lightgray}{-0.0023}&0.0024&0.0322 && \cellcolor{lightgray}{0.1203}&0.1375&0.1657\\
& 2.2385&\cellcolor{lightgray}{-0.0013}&-0.0131&0.0454 & & \cellcolor{lightgray}{0.1172}&0.1541&0.1568\\
& 2.537&-0.0674&-0.0464&\cellcolor{lightgray}{-0.0038} & & 0.2582&0.2428&\cellcolor{lightgray}{0.2283}\\
& 2.8355&-0.5926&\cellcolor{lightgray}{-0.5035}&-0.5179 & &\cellcolor{lightgray}{0.7421}&0.9565&0.9687\\ \hline 

$\theta$ & 1.098&\cellcolor{lightgray}{0.0024}&0.0325&0.0252 && \cellcolor{lightgray}{0.0620}&0.0668&0.0669\\
& 1.297&\cellcolor{lightgray}{0.0022}&0.0214&0.0140 && \cellcolor{lightgray}{0.0705}&0.0777&0.0792\\
& 1.495&\cellcolor{lightgray}{0.0034}&0.0169&0.0137 &&\cellcolor{lightgray}{0.0792}&0.0884&0.0914\\
& 1.693&\cellcolor{lightgray}{0.0048}&0.0119&0.0116 &&\cellcolor{lightgray}{0.0907}&0.0984&0.1046\\
& 1.891&0.0199&\cellcolor{lightgray}{0.0114}&0.0150 && 0.1306&\cellcolor{lightgray}{0.1105}&0.1194\\
& 2.089&0.0527&\cellcolor{lightgray}{0.0125}&0.0182 && 0.1703&\cellcolor{lightgray}{0.1205}&0.1309\\
& 2.287&0.0842&\cellcolor{lightgray}{0.0164}&0.0210 && 0.1986&\cellcolor{lightgray}{0.1340}&0.1463\\
& 2.485&0.0953&\cellcolor{lightgray}{0.0178}&0.0212 && 0.2142&\cellcolor{lightgray}{0.1394}&0.1524\\
& 2.683&0.1151&\cellcolor{lightgray}{0.0060}&0.0078 && 0.2350&\cellcolor{lightgray}{0.1351}&0.1489\\
& 2.881&0.1238&-0.0342&\cellcolor{lightgray}{-0.0258} && 0.2495&\cellcolor{lightgray}{0.1392}&0.1675\\ \hline

$\lambda$ & 0.32&0.0789&\cellcolor{lightgray}{0.0169}&0.0352 && 0.2219&\cellcolor{lightgray}{0.0558}&0.0582\\
& 0.3595&0.0735&\cellcolor{lightgray}{0.0121}&0.0315 && 0.2104&\cellcolor{lightgray}{0.0595}&0.0629\\
& 0.399&0.0675&\cellcolor{lightgray}{0.0084}&0.0283 && 0.1998&\cellcolor{lightgray}{0.0645}&0.0688\\
& 0.4385&0.0623&\cellcolor{lightgray}{0.0047}&0.0255 && 0.1894&\cellcolor{lightgray}{0.0721}&0.0746\\
& 0.478&0.0574&\cellcolor{lightgray}{0.0021}&0.0226 && 0.1791&\cellcolor{lightgray}{0.0772}&0.0790\\
& 0.518&0.0534&\cellcolor{lightgray}{0.0000}&0.0212 && 0.1705&\cellcolor{lightgray}{0.0808}&0.0824\\
& 0.5575&0.0464&\cellcolor{lightgray}{-0.0054}&0.0137 && 0.1617&\cellcolor{lightgray}{0.0855}&0.0866\\
& 0.597&0.0412&-0.0096&\cellcolor{lightgray}{0.0095} && 0.1535&\cellcolor{lightgray}{0.0898}&0.0905\\
& 0.6365&0.0364&-0.0161&\cellcolor{lightgray}{0.0032} && 0.1469&\cellcolor{lightgray}{0.0929}&0.0951\\
& 0.676&0.0297&-0.0224&\cellcolor{lightgray}{-0.0034} && 0.1416&\cellcolor{lightgray}{0.0985}&0.1009\\ \hline \hline

\end{tabular}
    \caption{Bias and standard deviation  of parameter estimates obtained by NF, NV and ML methods. Each row corresponds to a specific bin of the respective parameter. {Grey cells indicate the method that performed best in each case.}}
    \label{tab:summary}
\end{table}

\subsection{Real Data Illustration}

\begin{figure}
    \centering        \includegraphics[width=0.7\textwidth]{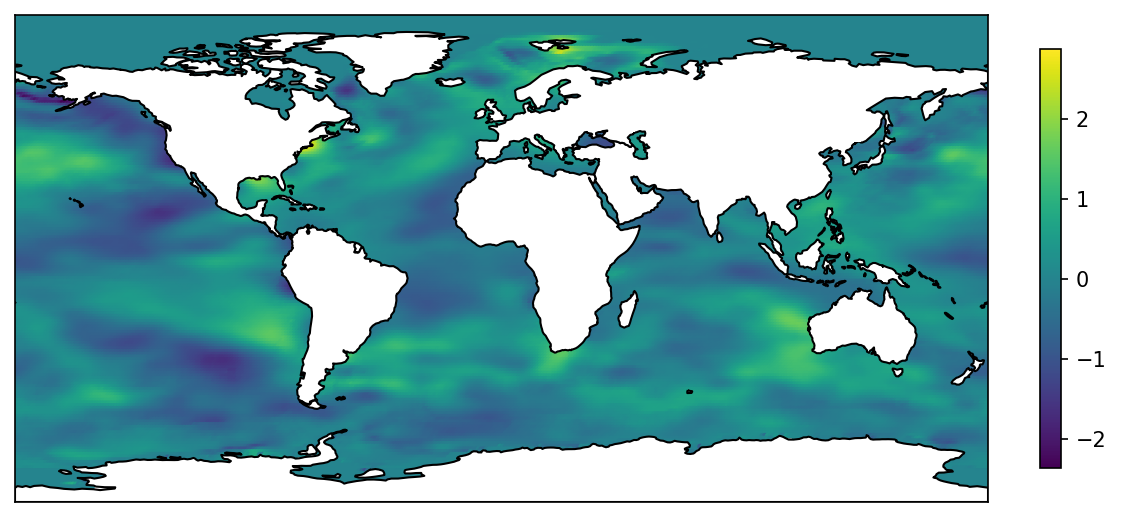}
    \caption{Anomalies of sea surface temperatures for March 2012.}
    \label{fig:sst}
\end{figure}

We now analyze the COBE Sea Surface Temperature data from NOAA PSL in Boulder, Colorado, available at \url{https://psl.noaa.gov} (see \citealp{ishii2005objective}).  {This dataset contains essential information for performing comprehensive global analyses of sea surface temperatures.} It  has a spatial resolution of $1^\circ \times 1^\circ$, which amounts to $27,761$ spatial locations. In our analysis, we focus on the mean anomalies for March 2012 (Figure \ref{fig:sst}).

Given that a global-scale process is unlikely to be stationary, the estimation methodologies are applied to different zones of the planet, creating a map that shows how parameters vary across different regions. Whether used directly for climatological investigations or as a starting point for developing versatile non-stationary geostatistical models, these surfaces hold significant value.  Although we analyze a different dataset, this approach is inspired by the study conducted by \cite{gerber2021fast}. Furthermore, we anticipate more informative results due to the incorporation of anisotropy.

%This can be valuable for understanding the local variations in the sea surface temperature spatial dependence structure. 
Specifically, we consider \(16 \times 16\) regions centered at each pixel and, after a standardization of the data inside each region, provide estimates of the parameters through NF, NV and ML. We only include pixels where a \(16 \times 16\) region can be created, meaning that pixels near the boundaries of the image or close to  land are excluded.

\begin{figure}
    \centering
    \includegraphics[width=0.45\linewidth]{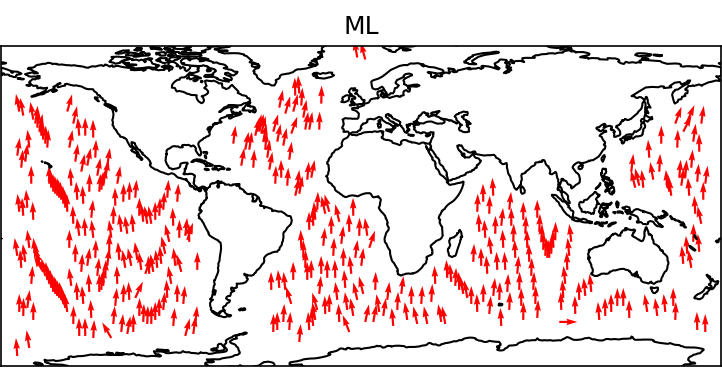}
     \includegraphics[width=0.45\linewidth]{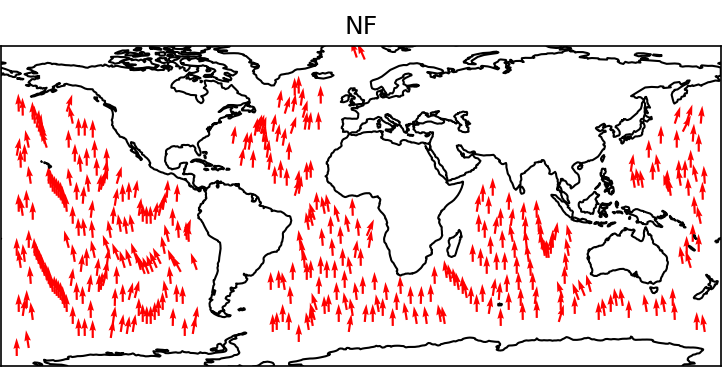}
      \includegraphics[width=0.45\linewidth]{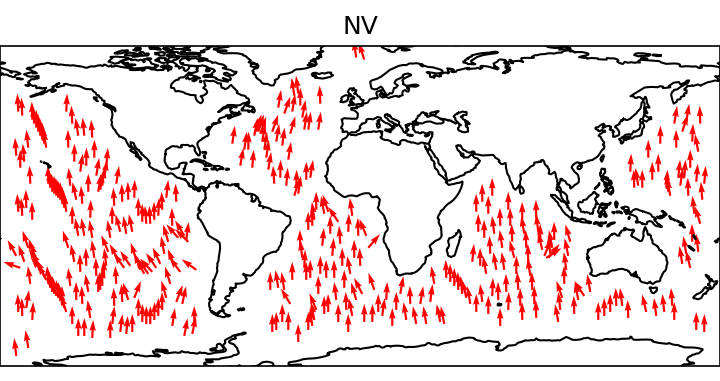}
    \caption{Estimates for $\alpha$ obtained from ML, NF and NV methods.}
    \label{fig:sst-alpha}
\end{figure}

Figure \ref{fig:sst-alpha} displays the results for the angular parameter ($\alpha$). The estimates are represented by arrows, and we only report the estimates at 390 pixels to enhance visualization. All methods deliver similar trends, demonstrating the effectiveness of the proposed approaches. {The local fluctuations of this parameter are influenced by ocean currents, atmospheric conditions and solar radiation, among others. }

\begin{figure}
    \centering
\includegraphics[width=\textwidth]{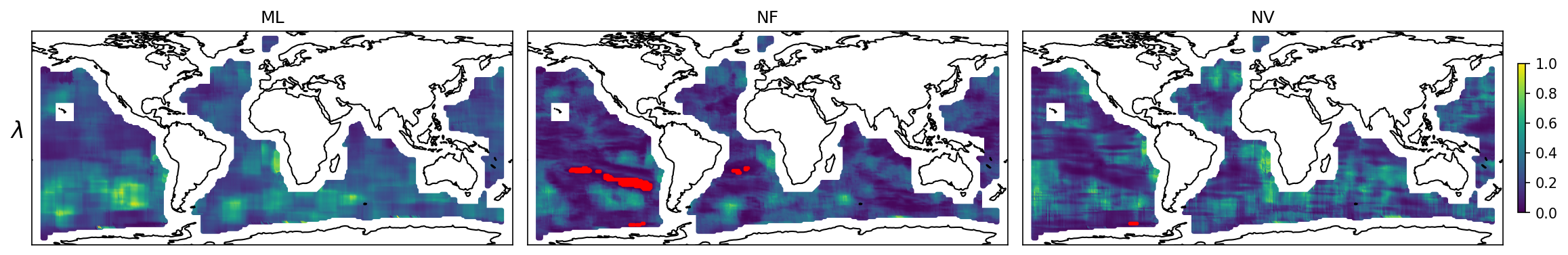}  
        \includegraphics[width=\textwidth]{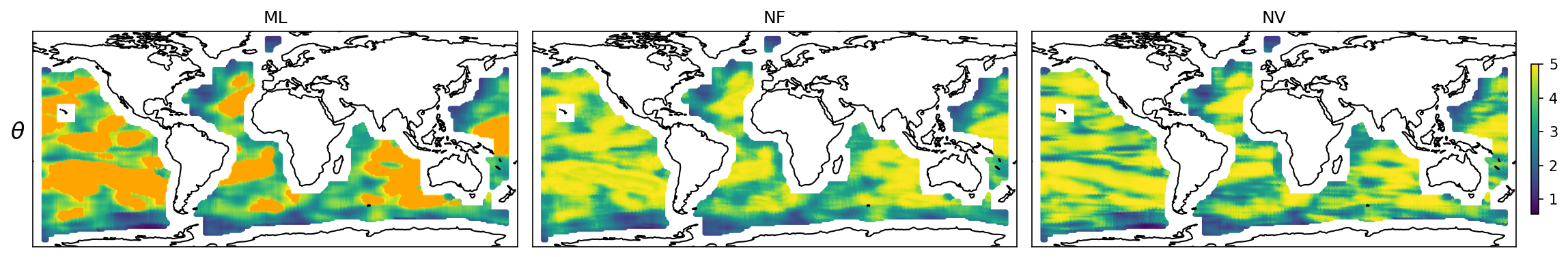}
    \caption{Estimates for $\lambda$ and $\theta$ obtained from ML, NF and NV methods.  Red pixels indicate estimates $\widehat{\lambda}$ outside the parameter space. Orange pixels represent estimates $\widehat{\theta}$ outside the domain of the test parameter grid.
}
    \label{fig:sst_results}
\end{figure}

\begin{figure}
    \centering
    \includegraphics[width=0.45\linewidth]{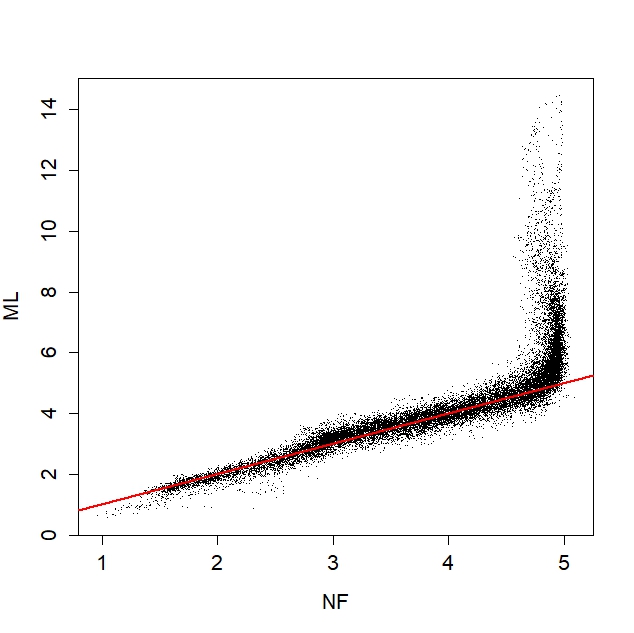}
\includegraphics[width=0.45\linewidth]{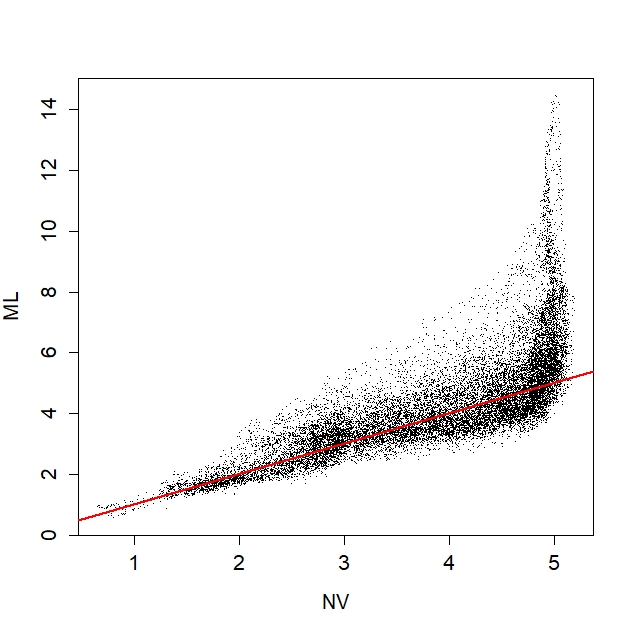}
    \caption{Scatter plots of the ML estimates against neural network estimates for  $\theta$.}
    \label{fig:scatter}
\end{figure}

The estimates for the anisotropy ratio ($\lambda$) and the correlation range ($\theta$) are presented in Figure \ref{fig:sst_results}. We observe similar patterns of the ratio across different methods, with large values in the Pacific Ocean in the southern hemisphere. It is important to note that NF and NV exhibit a small portion of estimates falling outside the parameter space (represented by red pixels). Specifically, these are negative values very close to zero. This is not surprising, as the networks were less trained at the boundaries of the interval $[0,1]$, making behaviors close to zonal anisotropy more difficult to identify. For the range parameter, we observe an overall consistent pattern across all methods. However, certain zones show ML estimates outside the test parameter domain (indicated by orange pixels). Some of these ML estimates exceed $10$, implying practical ranges greater than $45$, which is excessive for a $16 \times 16$ region.   As discussed in the previous section, we recall that an increase in the range parameter implies a strengthening of spatial dependence, leading to ML estimates with greater dispersion.   
In contrast, the neural network estimates do not exhibit this behavior, as they have been trained within a specific interval of coherent values. Figure \ref{fig:scatter} reinforces this by
 presenting scatter plots comparing ML against NF and NV methods, illustrating that the methods align correctly except when the range parameter approaches the upper bound of the test parameter grid.

In terms of computational speed, the neural network-based methods are significantly faster, with NF taking only 41 seconds and NV just 11 seconds to complete the task, compared to the 3 hours required by the ML method.

\section{Discussion and Concluding Remarks}
\label{discussion}

This paper introduces a methodology based on neural networks for parameter estimation in geometrically anisotropic Gaussian random fields {in the plane}. We have demonstrated the effectiveness of our approach through simulation studies and an application to sea surface temperatures. Beyond achieving statistical accuracy comparable to traditional ML methods, our neural network techniques exhibit notable robustness. For instance, they enable precise identification of the angle of anisotropy, exhibit a more gradual increase in variability of range estimates as the range parameter grows, and prevent ratio estimates from degenerating to extreme values within the ratio interval. In this regard, this work complements recent studies in the field.

In many disciplines of the earth and environmental sciences and natural resources engineering, spatial data often exhibit distinctive characteristics, in particular correlation functions with a linear behavior near the origin and/or short-scale variability (nugget effect) caused by  measurement errors \citep{chiles2012geostatistics}. Conventional ML estimation procedures address these characteristics by fitting an appropriate covariance model from a broad range of options. However, both NF and NV would necessitate training a different network, as the networks presented in the previous sections were specifically calibrated for variograms with a parabolic shape near the origin and no nugget effect, focusing on achieving versatile geometric anisotropies. Although we have omitted these variants for brevity, we anticipate that the architectures proposed in this paper will serve as effective building blocks for adapting to these scenarios.

%\textcolor{red}{Also, 3D data: does it make some difference? I guess that it introduces more parameters in the ML estimation (three ranges and three angles for geometric anisotropy): does the procedure remain applicable?}

At the same time, we have identified that using covariance models with a very large number of parameters results in significant computational costs during network training, as it requires covering a large region of the parameter space. Consequently, the number of scenarios needed increases dramatically with the dimensionality of the parameter vector. This could be particularly challenging when dealing with spatio-temporal or vector-valued random fields.  {Geometric anisotropy in three dimensions also presents a significant challenge, as the number of ranges increases to three, and the number of angles also expands to three.}   {We thus believe that designing innovative neural network approaches to handle a large number of parameters is an interesting direction to explore.}

The applicability of these techniques is virtually limitless, as most geostatistical applications involve models where evaluating exact likelihood and assessing unbiased predictions of minimum variance present significant computational and analytical challenges.

\section*{Acknowledgments}

Alejandro Villaz\'on was funded by ANID BECAS/MAGÍSTER NACIONAL 22240803 and by Programa de Incentivo a la Iniciación Científica (PIIC) 011/2024,  Dirección de Postgrado, Universidad Técnica Federico Santa María. 
Alfredo Alegr\'ia acknowledges the funding of Universidad Técnica Federico Santa María, through grant Proyectos Internos USM 2023 PI$_{-}$LIR$_{-}$23$_{-}$11. 
Xavier Emery acknowledges the funding of the National Agency for Research and Development of Chile, through grant ANID PIA AFB230001.

\bibliography{mybib}
\bibliographystyle{apalike}

\end{document}